\pdfoutput=1
\documentclass[lettersize,journal]{IEEEtran}
\usepackage{amsmath,amsfonts,amssymb}
\usepackage{amsthm}
\usepackage{algorithmic}
\usepackage{algorithm}
\usepackage{array}
\usepackage{textcomp}
\usepackage{stfloats}
\usepackage{url}
\usepackage{verbatim}
\usepackage{graphicx}  
\usepackage{cite}
\usepackage{float}
\usepackage{caption}
\usepackage{subcaption}
\usepackage{wrapfig}    

\usepackage[colorlinks=true,
            linkcolor=red,    
            citecolor=green,  
            urlcolor=blue     
           ]{hyperref}

\usepackage{orcidlink}  

\begin{document}

\title{Fault Detection and Human Intervention in Vehicle Platooning:\\A Multi-Model Framework}

\author{Farid Mafi\textsuperscript{1}\orcidlink{0000-0002-3726-0282},  
        Mohammad Pirani\textsuperscript{2}\orcidlink{0000-0003-2677-2140}, Senior Member, IEEE
\thanks{\textsuperscript{1}F. Mafi is with the Department of Mechanical and Mechatronics Engineering, University of Waterloo, Waterloo, ON N2L 3G1, Canada (e-mail: fmafisho@waterloo.ca).}
\thanks{\textsuperscript{2}M. Pirani is with the Department of Mechanical Engineering, University of Ottawa, Ottawa, ON K1N 6N5, Canada (e-mail: mpirani@uottawa.ca).}
}



\maketitle

\begin{abstract}
Vehicle platooning has been a promising solution for improving traffic efficiency and throughput. However, a failure in a single vehicle, including communication loss with neighboring vehicles, can significantly disrupt platoon performance and potentially trigger cascading effects. Similar to modern autonomous vehicles, platoon systems require human drivers to take control during failures, leading to scenarios where vehicles are operated by drivers with diverse driving styles. This paper presents a novel multi-model approach for simultaneously identifying signal drop locations and driver attitudes in vehicular platoons using only tail vehicle measurements. The proposed method distinguishes between attentive and distracted driver behaviors by analyzing the propagation patterns of disturbances through the platoon system. Beyond its application in platooning, our methodology for detecting driver behavior using a multi-model approach provides a novel framework for human driver identification. To enhance computational efficiency for real-time applications, we introduce a blending-based identification method utilizing chosen models and weighted interpolation, significantly reducing the number of required models while maintaining detection accuracy. The effectiveness of our approach is validated through high-fidelity CarSim/Simulink environment simulations. Results demonstrate that the proposed method can accurately identify both the location of signal drops and the corresponding driver behavior. This approach minimizes the complexity and cost of fault detection while ensuring accuracy and reliability.
\end{abstract}


\IEEEpubidadjcol

\section{Introduction}

\subsection{Motivation}
\IEEEPARstart{V}{ehicle} platooning has emerged as a valuable topology for improving transportation efficiency, fuel consumption, and road capacity utilization, with particularly widespread adoption in heavy-duty vehicles and freight transportation systems. By maintaining reduced inter-vehicle distances through coordinated control, platoons can achieve significant reductions in aerodynamic drag and fuel consumption. However, these formations are inherently vulnerable to various types of failures, ranging from communication disruptions and sensor malfunctions to deliberate adversarial actions, each potentially leading to collision risks and substantial economic losses. When such failures occur in commercial platooning operations, human intervention becomes necessary as a critical safety measure, requiring drivers to take immediate control of affected vehicles. In these scenarios, understanding the driver's behavior and attention state is crucial for maintaining overall platoon stability and ensuring safety. The variability in human response—whether attentive or distracted—significantly impacts how disturbances propagate through the platoon system, directly affecting the efficacy of safety protocols. This research addresses the dual challenge of detecting communication failures while simultaneously identifying driver behavior patterns using minimal sensor information, providing a practical solution for enhancing platoon safety and reliability in real-world implementations.

\subsection{Literature Review}
The coordination and control of multi-vehicle systems has been an active area of research in intelligent transportation, with significant theoretical and practical developments over the past two decades \cite{turri2017, li2015}. The formation of automated vehicle groups operating in close proximity presents unique challenges in control architecture, communication reliability, and safety assurance \cite{pirani2022stable, zheng2016}. These challenges have prompted extensive research across various domains, from string stability theory \cite{swaroop1996} to human-machine interaction \cite{jin2014}. String stability, which prevents the amplification of disturbances along the vehicle string, has been extensively studied using various approaches \cite{feng2019}. Recent advances in connected and automated vehicle technology have enabled the implementation of sophisticated control strategies that address both safety and performance objectives \cite{alam2015, besselink2017}.

Building upon these foundational concepts, researchers have explored different control architectures for platoon systems. Traditional methods focused on predecessor-following and bidirectional control architectures \cite{hao2013}, with each architecture presenting distinct advantages and limitations regarding stability margins and disturbance propagation. Recent studies have demonstrated that the predecessor-following architecture, while simpler to implement, exhibits exponential scaling of disturbance amplification with platoon size \cite{seiler2004}. In contrast, bidirectional architectures show polynomial scaling, offering better robustness but requiring more complex control strategies \cite{zheng2016}.

In the realm of practical implementation, communication reliability remains a critical challenge in platoon systems. Signal drops can significantly impact platoon stability and safety \cite{liu2001}, with research showing that communication delays and packet losses can lead to string instability and potentially dangerous situations \cite{oncu2012}. Various approaches have been proposed to address these issues, including graceful degradation strategies \cite{ploeg2015} and delay-based spacing policies \cite{besselink2017}. Additionally, the integration of human factors has gained increasing attention, particularly in mixed traffic scenarios where automated and human-driven vehicles coexist \cite{mahbub2021}. Studies have identified two primary driver modes: attentive and distracted \cite{jin2014}, which significantly influence the response characteristics and stability of the platoon system. Driver attention levels can be classified through various indicators, including reaction times and control input patterns \cite{darbha2005}, and their impact on string stability has been analyzed using both theoretical and experimental approaches \cite{alam2015}.

The evolution of fault detection methods has led to significant advances in platoon system monitoring. While traditional methods primarily focused on direct measurement of communication signals and vehicle states \cite{fernandes2012}, these approaches often require extensive sensor networks and may not be practical in real-world scenarios. Multi-model approaches for fault detection have emerged as a promising alternative \cite{shaw2007}, capable of identifying both system faults and driver behavior patterns by analyzing the propagation of disturbances through the platoon \cite{zheng2016}. Recent studies have demonstrated the effectiveness of using only tail vehicle measurements for fault detection and classification \cite{ploeg2014}. Advanced control strategies have been developed to maintain platoon stability under various uncertainty conditions \cite{li2017}, including robust control methods \cite{gao2018}, adaptive approaches \cite{zheng2018}, and learning-based techniques \cite{jin2017}. The integration of multiple control objectives, including string stability, fuel efficiency, and fault tolerance, has become a key research direction \cite{turri2017}.

Recent research has focused on developing comprehensive frameworks that can handle both communication failures and driver behavior variations \cite{besselink2018}, often employing hybrid approaches that combine model-based and data-driven methods to achieve robust performance \cite{feng2019}. However, several challenges remain unresolved in the field. The simultaneous identification of communication failures and driver behavior patterns using limited measurements represents a significant research gap \cite{mahbub2021}. Additionally, the development of efficient fault detection methods that can operate under realistic conditions with limited sensor information requires further investigation \cite{feng2019}. The proposed multi-model approach for identifying signal drops and driver attitudes using only tail vehicle measurements addresses these gaps while providing a practical solution for real-world implementation, building upon existing work in fault detection and driver behavior classification while offering new insights into platoon system monitoring and control.

\subsection{Contributions}
The main contributions of this paper can be summarized as follows:

\begin{enumerate}
    \item  We propose a multi-model framework for detecting and identifying signal drops in two vehicular platoon topologies: Predecessor-Following and Bidirectional. This framework relies solely on the dynamic response of the last vehicle, eliminating the need for extensive sensor networks across the platoon.
    
    \item Using the multi-model scheme, we identify the driver's attention states (attentive/distracted) when taking over the vehicle during failures. This extends the theoretical understanding of how human factors influence platoon dynamics in fault scenarios.

    \item We introduce a blending-based multi-model identification method that significantly reduces computational complexity, enhancing real-time applicability while preserving detection accuracy.
\end{enumerate}

Beyond its applications in platooning systems, our multi-model driver behavior inference methodology contributes to the broader literature on human behavior detection in autonomous driving, where approaches have been predominantly data-driven.

\section{Problem Formulation}

\subsection{System Dynamics}
\begin{figure}[!t]
	\centering
	\includegraphics[width=\columnwidth]{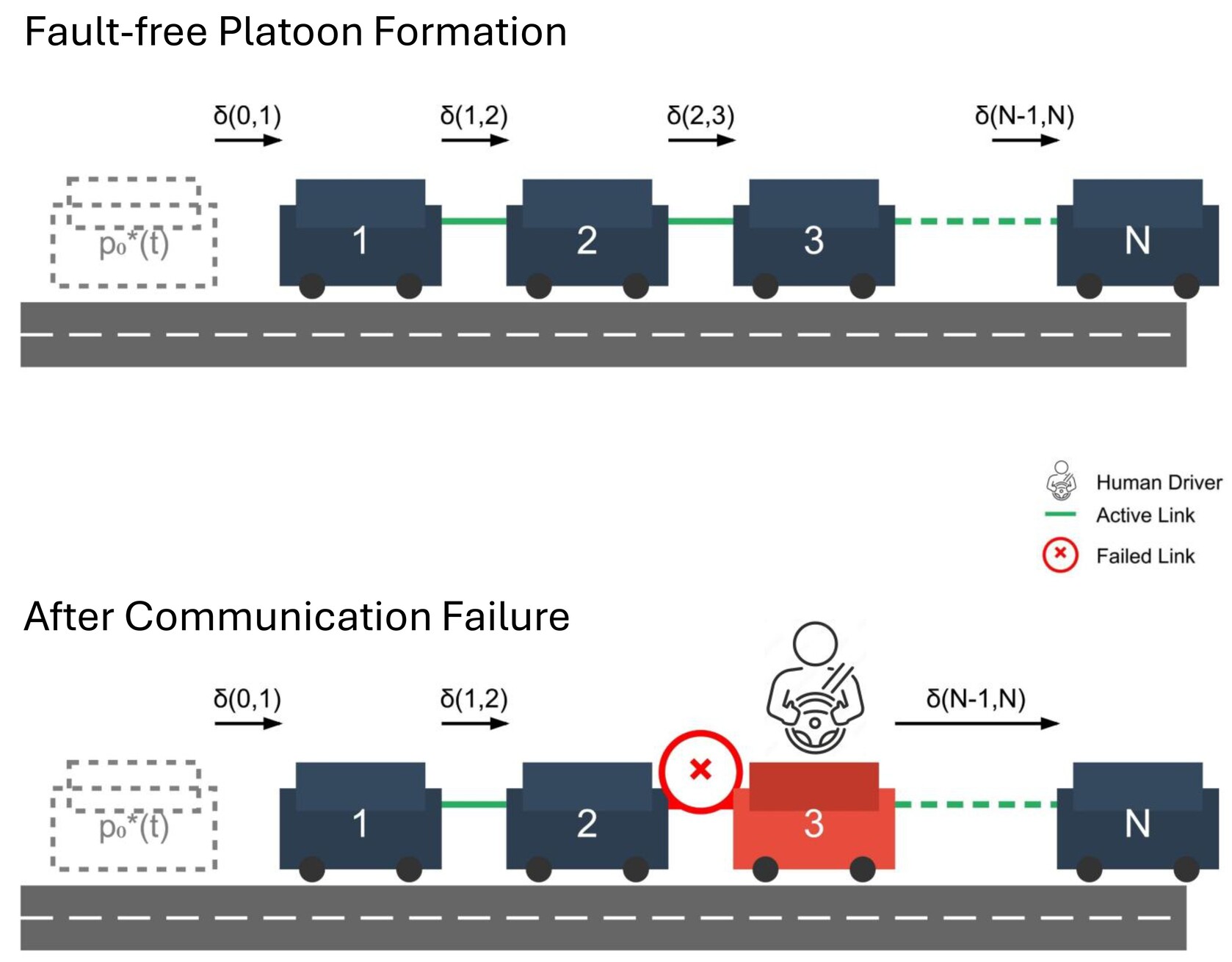}
    \caption{One-dimensional platoon formation with a virtual reference vehicle and N following vehicles: a) Initial Formation b) After Failure.}
    \label{fig:platoon}
\end{figure}

Consider a platoon of $N$ homogeneous vehicles moving in a single-lane straight path, as shown in Fig. \ref{fig:platoon}. Each vehicle is modeled as a double integrator \cite{hao2013}:
\begin{equation}\label{eq1}
    \ddot{p}_i = u_i + w_i, \quad i \in \{1,2,\ldots,N\}.
\end{equation}

\noindent where $p_i$ denotes the position of the $i$th vehicle, $u_i$ is the control input, and $w_i$ represents external disturbances. The control objective is to maintain a rigid formation geometry while following a desired trajectory. The desired geometry is specified by inter-vehicular gaps $\delta_{(i-1,i)}$ for $i \in \{1,\ldots,N\}$, where $\delta_{(i-1,i)}$ represents the desired distance between vehicles $i$ and $i-1$. These gaps must satisfy mutual consistency: $\delta_{(i,k)} = \delta_{(i,j)} + \delta_{(j,k)}$ for every triple $(i,j,k)$ where $i \leq j \leq k$. The desired trajectory is provided through a fictitious reference vehicle (index 0) with trajectory $p_0^*(t)$. Only the first vehicle has access to this reference trajectory. The desired trajectory for the $i$th vehicle is then:
\begin{equation}\label{eq2}
    p_i^*(t) = p_0^*(t) - \delta_{(0,i)} = p_0^*(t) - \sum_{j=1}^i \delta_{(j-1,j)}.
\end{equation}

\noindent\textbf{Predecessor-Following Architecture:}
In this architecture, each vehicle's control action depends only on relative measurements from its immediate front neighbor:
\begin{equation}\label{eq4}
    u_i = f(p_i - p_{i-1} + \delta_{(i-1,i)}) - g(\dot{p}_i - \dot{p}_{i-1}).
\end{equation}

\noindent where $f,g: \mathbb{R} \rightarrow \mathbb{R}$ are scalar functions that can be either linear or nonlinear.

\noindent\textbf{Bidirectional Architecture:}
In this architecture, each vehicle's control action depends equally on relative measurements from both its immediate front and back neighbors:
\begin{equation}\label{eq5}
\begin{split}
    u_i = & f(p_i - p_{i-1} + \delta_{(i-1,i)}) - g(\dot{p}_i - \dot{p}_{i-1}) \\
    & - f(p_i - p_{i+1} - \delta_{(i,i+1)}) - g(\dot{p}_i - \dot{p}_{i+1}),
\end{split}
\end{equation}

\noindent for $i \in \{1,\ldots,N-1\}$, and
\begin{equation}\label{eq6}
    u_N = f(p_N - p_{N-1} + \delta_{(N-1,N)}) - g(\dot{p}_N - \dot{p}_{N-1}).
\end{equation}

For vehicles moving in formation, communication reliability is crucial. When failures occur, the system must transition seamlessly to human control while maintaining safety. In particular, when a communication fault occurs at the $k$-th vehicle at time $t_f$, only this vehicle is aware of the connection loss with its predecessor. The fault status between adjacent vehicles $i$ and $j$ is characterized by:
\begin{equation}
    \xi_{ij}(t) = \begin{cases}
        1, & \text{if communication link is active} \\
        0, & \text{if communication link has failed}
    \end{cases}
\end{equation}

\noindent where the fault detection time $t_f$ is defined as:
\begin{equation}
    t_f = \inf\{t > 0: \xi_{k,k-1}(t) = 0\}.
\end{equation}

Upon fault detection at vehicle $k$, control transitions from autonomous to human-driven mode according to Fig. \ref{fig:platoon}, with the driver implementing an immediate safety protocol through controlled deceleration. Here, $u_{\rm auto,k}(t)$ represents the nominal autonomous control law defined in equation \eqref{eq4} for the predecessor-following architecture or equation \eqref{eq5} for the bidirectional architecture. This human-in-the-loop intervention is modeled as:

\begin{equation}
    u_k(t) = \begin{cases}
        u_{\rm auto,k}(t), & t < t_f \\
        u_{\rm drv,k}(t) - a_{\rm saf}, & t \geq t_f
    \end{cases}
\end{equation}

\noindent where $a_{\rm saf}$ represents a constant safe deceleration rate initiated by the driver. The driver's base control input $u_{\rm drv,k}(t)$ follows either the attentive or distracted model:

\begin{equation}
    u_{\rm drv,k}(t) = G_h(s)[\dot{p}_{k-1}(t) - \dot{p}_k(t)],
\end{equation}

\noindent where $G_h(s)$ represents the driver transfer function which will be defined later and is related to the driver's behavior and driving mode. The implementation of this driver-controlled deceleration at vehicle $k$ initiates a chain reaction in the platoon. This cascading response creates a natural deceleration wave through the platoon, where following vehicles adjust their velocities in response to the $k$-th vehicle's safety protocol without explicit knowledge of the fault. The deceleration initiated by the driver at vehicle $k$ propagates through the formation as each following vehicle responds to the changing distance from its predecessor.

\subsection{Multi-Model Identification Framework}

Consider a platoon of $N$ vehicles where a communication failure can occur at any vehicle $k \in \{1,\ldots,N\}$, and the driver at vehicle $k$ can be in either an attentive or distracted state. Let $\mathcal{M}$ denote the set of all possible model combinations:

\begin{equation}
    \mathcal{M} = \{M_{k,d}: k \in \{1,\ldots,N\}, d \in \{A,D\}\}.
\end{equation}

\noindent where $A$ and $D$ represent attentive and distracted driver states, respectively. Each model $M_{k,d}$ corresponds to a specific combination of fault location $k$ and driver type $d$. Among these possible models, the one that produces the minimum identification error between the predicted and actual system outputs will be selected as the true model. The indices $(k,d)$ of this best-matching model directly identify both the location of the signal drop and the type of driver behavior present in the system.

The fundamental problem addressed in this paper is to both identify the location of signal drops and the driver's behavior type (attentive or distracted) in vehicular platoons using only measurements from the tail vehicle. The proposed multi-model approach aims to detect these parameters in platoons operating under either predecessor-following or symmetric bidirectional control architectures. When a communication fault occurs between vehicles, the affected vehicle transitions to human control, with the driver's attention state impacting disturbance propagation through the system. Once the fault location and driver type are identified through the proposed multi-model approach, appropriate safety protocols can be implemented to maintain platoon stability and ensure safe operation.

\section{The Proposed Fault Detection Approach}
This section presents our fault detection methodology for vehicle platoons. We first detail the multi-model identification framework, followed by the derivation of transfer functions for different platoon architectures and driver types. These transfer functions are then combined to create models representing specific fault scenarios, enabling simultaneous detection of signal drop location and driver behavior through tail vehicle analysis.

\subsection{Multi-Model Detection Method}

\begin{figure}[!t]
    \centering
    \includegraphics[width=\columnwidth]{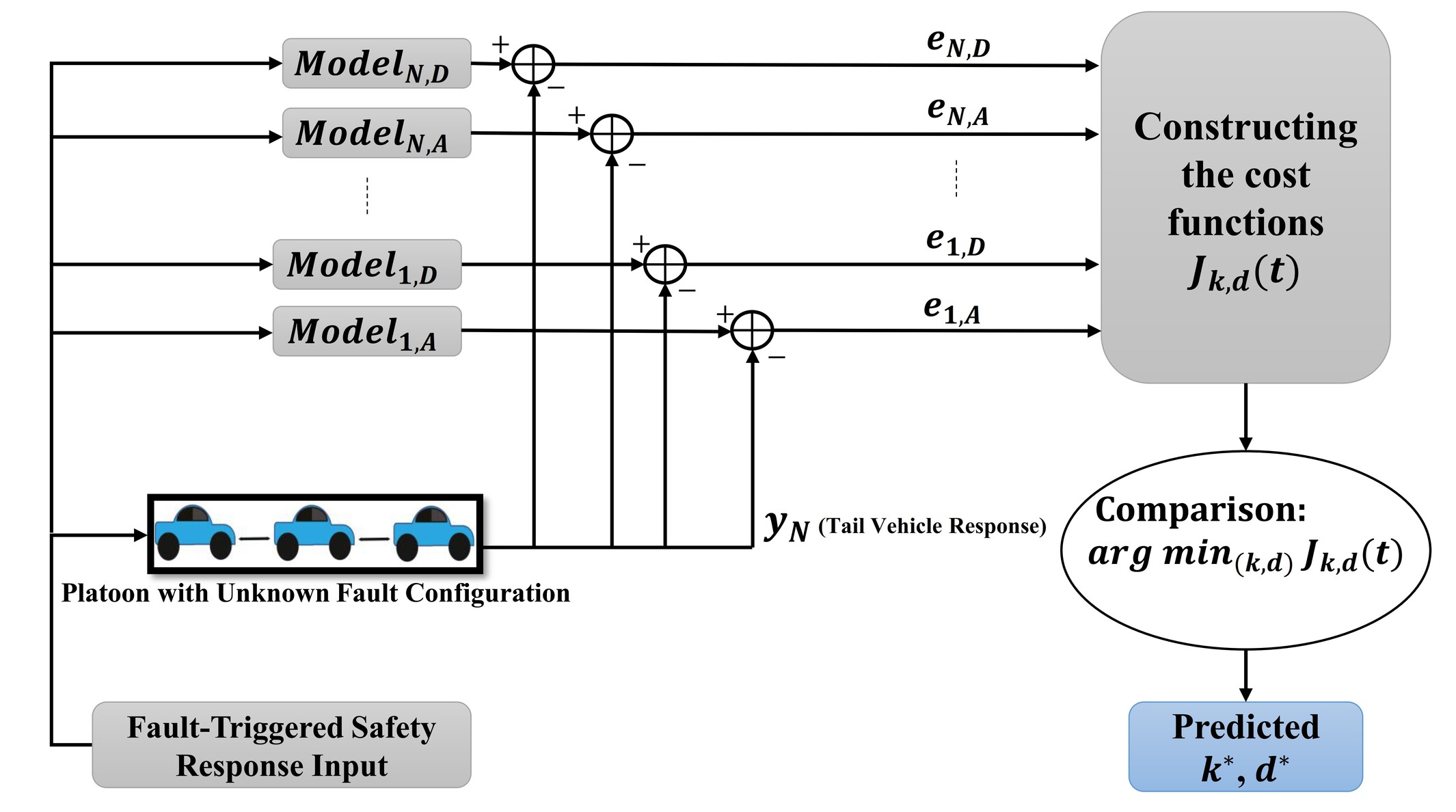}
    \caption{Multi-model scheme for simultaneous identification of signal drop location and driver behavior in vehicular platoons}
    \label{fig:scheme}
\end{figure}

To implement the proposed identification framework, we need to establish a systematic method for evaluating and comparing different models. This is achieved through a careful analysis of the system outputs and the formulation of appropriate error metrics. For each model $M_{k,d}$, the combined transfer function incorporating both vehicle dynamics and human driver behavior is considered. As illustrated in Fig. \ref{fig:scheme}, the multi-model scheme comprises a set of fixed models designed to cover all possible fault scenarios in the platoon system. Each designed model consists of fixed parameters that correspond to specific combinations of fault location and driver behavior type. These fixed models are distributed to encompass the complete range of fault possibilities, ensuring comprehensive coverage of all potential signal drop locations and driver response patterns.

At time $t$, let $y_N(t)$ represent the measured output from the last vehicle in the platoon, and let $\hat{y}_{k,d}(t)$ denote the predicted output generated by model $M_{k,d}$. The identification error between the actual and predicted outputs is then defined as:

\begin{equation}
    e_{k,d}(t) = y_N(t) - \hat{y}_{k,d}(t).
\end{equation}

To evaluate the performance of each model, we define a cost function $J_{k,d}(t)$ that combines both instantaneous and historical error information. This weighted cost function with forgetting factor takes the form:

\begin{equation}
    J_{k,d}(t) = \alpha\|e_{k,d}(t)\|^2 + \beta \int_{t_f}^t e^{-\lambda(t-\tau)}\|e_{k,d}(\tau)\|^2 d\tau.
    \label{eq:cost_function}
\end{equation}

\noindent The structure of this cost function incorporates several key parameters:
\begin{itemize}
    \item $\alpha > 0$ weights the contribution of instantaneous error
    \item $\beta > 0$ weights the contribution of historical error
    \item $\lambda > 0$ serves as a forgetting factor that gradually reduces the influence of past errors
    \item $t_f$ marks the time when the fault occurs, serving as the lower bound of integration
\end{itemize}

Using this cost function, the true fault location $k^*$ and driver type $d^*$ can be identified through:

\begin{equation}
    (k^*, d^*) = \arg\min_{(k,d)} J_{k,d}(t).
\end{equation}

Through this optimization, the model with minimum cost function value reveals both the location of the signal drop $k^*$ and the corresponding driver behavior type $d^*$, thereby completing the fault identification process. This simultaneous identification enables appropriate safety measures to be implemented based on both the fault location and driver characteristics.

\subsection{First-to-Last Transfer Function Analysis}
To analyze the propagation of disturbances through the platoon, we derive transfer functions from the first vehicle to the last vehicle for both control architectures. The position tracking error for each vehicle is defined as:

\begin{equation}\label{eq3}
    \tilde{p}_i \triangleq p_i - p_i^*.
\end{equation}

For the predecessor-following architecture, we derive a transfer function that characterizes how position errors propagate through the platoon chain. Through successive substitution and algebraic manipulation (see Appendix~\ref{appendix:transfer_derivation} for complete derivation), we obtain the first-to-last transfer function:

\begin{equation} \label{eq14}
    G_{\rm PF}(s) = \frac{\tilde{P}_N(s)}{\tilde{P}_1(s)}= \left(\frac{b_0 s + k_0}{s^2 + b_0 s + k_0}\right)^{N-1}.
\end{equation}

\noindent This transfer function reveals how disturbances can amplify through the platoon, with the exponent $(N-1)$ indicating potential string instability for larger platoon sizes.

For the symmetric bidirectional architecture, the analysis considers both forward and backward interactions between vehicles. The resulting transfer function, derived through matrix analysis of the coupled system (detailed derivation in Appendix~\ref{appendix:transfer_derivation}), takes the form:

\begin{equation} \label{eq24}
G_{\rm SB}(s) = \frac{\tilde{P}_N(s)}{\tilde{P}_1(s)} = \frac{(-1)^{N+1}\beta(s)^{N-1}}{D_N(s)},
\end{equation}

\noindent where the denominator polynomial $D_N(s)$ is given by:

\begin{align} \label{eq25}
D_N(s) = &(s^2 + b_0s + k_0)(s^2 + 2b_0s + 2k_0)^{N-1} \nonumber \\
&- (b_0s + k_0)^2(s^2 + 2b_0s + 2k_0)^{N-2}.
\end{align}

\noindent This more complex structure of the bidirectional transfer function reflects the enhanced stability properties achieved through bilateral coupling between vehicles. The polynomial structure of the denominator, in particular, indicates improved string stability characteristics compared to the predecessor-following architecture.

These transfer functions provide the foundation for our subsequent analysis of fault detection and driver behavior identification. They enable us to characterize how different types of disturbances propagate through the platoon under various control architectures and operating conditions.

\subsection{Transfer Function Analysis and Driver Behavior}
The behavior of human drivers in vehicle-following scenarios significantly influences platoon dynamics, particularly during communication failures. Based on experimental studies \cite{macadam2003understanding}, driver behavior can be characterized through transfer functions that capture key cognitive and motor response characteristics:

\begin{equation}
   G_h(s) = K \frac{1 + T_{\rm z} s}{1 + 2\gamma T_{\rm w} s + T_{\rm w}^2 s^2} e^{-T_{\rm d} s} = \frac{\dot{P}_h(s)}{\dot{P}_n(s)},
\end{equation}

\noindent where $K$ represents the steady-state velocity matching ratio, $T_{\rm z}$ characterizes anticipatory behavior, $T_{\rm w}$ denotes the natural frequency time constant, $\gamma$ represents the damping factor, and $T_{\rm d}$ indicates cognitive and neuromuscular delays. This model distinguishes between attentive and distracted drivers through their distinct response parameters:
\begin{alignat}{2}
   &H_{\rm att}(s) &&= \frac{1 + 5.41s}{1 + 2(0.54)(4.15)s + (4.15)^2 s^2} e^{-0.324s}, \\
   &H_{\rm dist}(s) &&= \frac{1 + 6.96s}{1 + 2(0.65)(4.76)s + (4.76)^2 s^2} e^{-0.512s}.
\end{alignat}

\noindent The key distinction between these driver types manifests in their response characteristics, with distracted drivers exhibiting longer delays (0.512s vs 0.324s) and more oscillatory behavior (damping factor 0.65 vs 0.54). These differences significantly impact platoon stability and performance during communication failures.

When these driver models are integrated with the platoon dynamics under different control architectures (predecessor-following or symmetric bidirectional), the resulting combined transfer functions capture how disturbances propagate through the system. Attentive drivers maintain better string stability due to their faster response times and precise control actions, while distracted drivers may introduce additional oscillations and potential instabilities. This fundamental difference in behavior patterns enables our fault detection methodology to accurately identify both the fault location and driver attention state using tail vehicle measurements.

\subsection{A Blending Multi-Model Approach}
In certain dynamic configurations, the system inherently requires a high number of models, causing computational demands to grow exponentially. This significant increase in identification models creates performance challenges that may limit real-time processing capabilities. To address this challenge, we propose a two-step identification procedure based on boundary models and weighted interpolation.

Instead of maintaining separate models for each possible platoon length, we utilize two boundary models $G_1(s)$ and $G_2(s)$ that represent the extremal cases of the uncertainty space:

\begin{equation}
    G_1(s) = \left(\frac{b_0s + k_0}{s^2 + b_0s + k_0}\right)^{N_1-1} H(s),
\end{equation}

\begin{equation}
    G_2(s) = \left(\frac{b_0s + k_0}{s^2 + b_0s + k_0}\right)^{N_2-1} H(s),
\end{equation}

\noindent where $N_1$ and $N_2$ represent the minimum and maximum possible platoon lengths after potential signal drops, respectively. While these equations represent the predecessor-following architecture, the same blending procedure can be applied to the symmetric bidirectional architecture by utilizing its corresponding transfer function structure. The predicted system response is then computed as a weighted sum of these boundary models:
\begin{equation}
    G_{\rm pred}(s) = W_1G_1(s) + W_2G_2(s),
\end{equation}

\noindent where weights $W_1$ and $W_2$ are determined through a moving-window least squares optimization subject to the constraints $W_1 + W_2 = 1$ and $W_1, W_2 \geq 0$. From these weights, we can determine an effective platoon length $N_{\rm eff}$ that characterizes the system's behavior (see Appendix~\ref{appendix:blending_derivation} for the complete derivation):
\begin{equation}\label{eq:blend}
    \begin{split}
    N_{\rm eff} = 1 + (N_2-N_1) & \frac{\ln\big(W_1(W_1/W_2)^{(N_1-1)/(N_2-N_1)} + }{} \\
    & \frac{W_2(W_1/W_2)^{(N_2-1)/(N_2-N_1)}\big)}{\ln(W_1/W_2)}.
    \end{split}
\end{equation}

\begin{figure}[!t]
	\centering
	\includegraphics[width=\columnwidth]{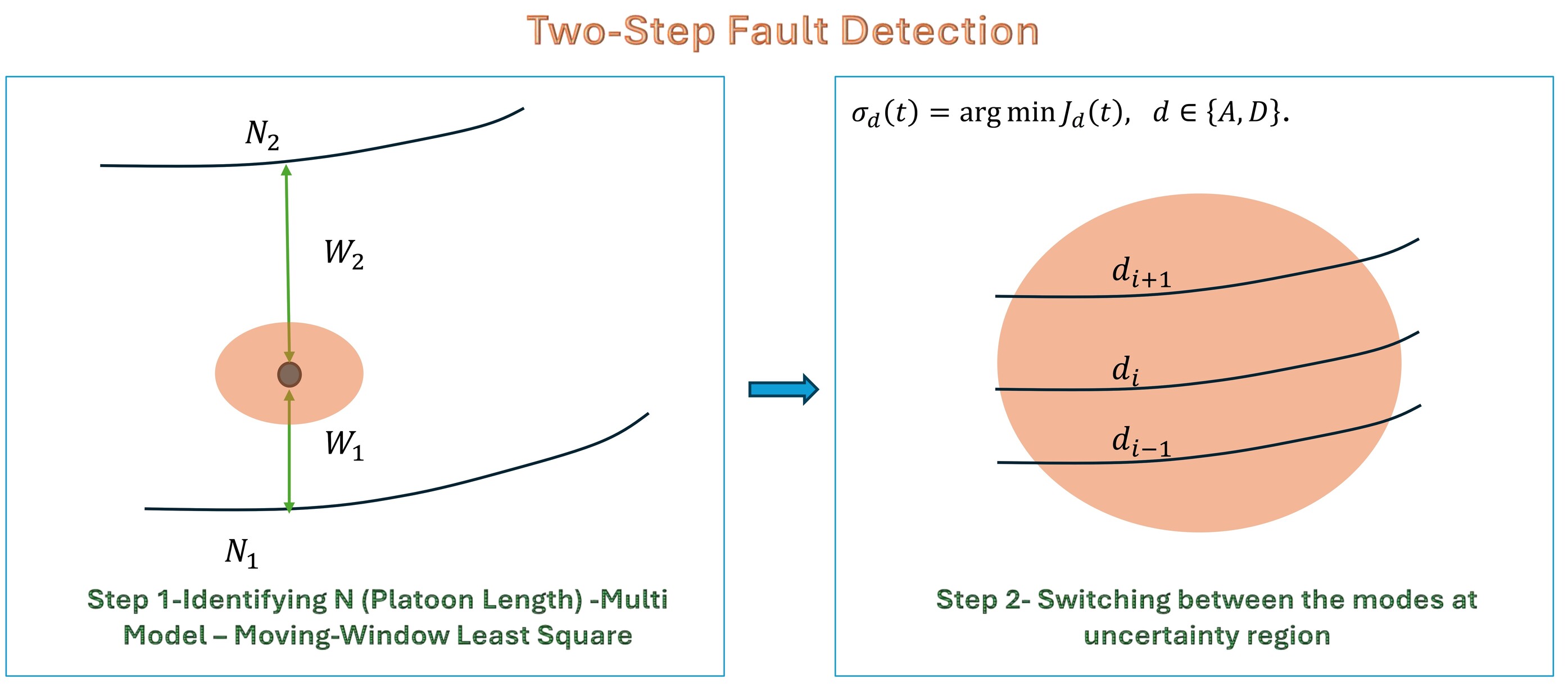}
	\caption{Proposed Blending-Based Fault Detection Approach.}
	\label{Blending_Architecture}
\end{figure}

With the completion of this first-step identification, the second step proceeds to identify the driver behavior type using the methodology outlined in the previous section. Specifically, the model exhibiting the smallest error metric is selected as the representative driver behavior. Fig.~\ref{Blending_Architecture} illustrates this two-step identification configuration.

This blending approach significantly reduces computational complexity while maintaining identification accuracy by capturing the system's behavior through continuous interpolation between boundary models.

\subsection{Uniqueness of the Identification Solution}

A critical consideration in multi-model identification frameworks is ensuring that the solution is unique—that is, there exists exactly one combination of fault location $k$ and driver type $d$ that minimizes the cost function $J_{k,d}(t)$. The uniqueness property guarantees that the identification process converges to the correct fault location and driver behavior, which is essential for implementing appropriate safety protocols.

The uniqueness of our approach is established through the monotonicity of the cost function with respect to the length of the platoon behind the faulty vehicle $(N-k)$. For the predecessor-following architecture, this monotonicity emerges from the exponential structure of the transfer function, while for the symmetric bidirectional architecture, it arises from the specific eigenvalue distribution of the system matrix.

For a platoon with actual fault at position $k^*$ with driver behavior $d^*$, we can demonstrate that:

\begin{equation}
    J_{k,d}(t) > J_{k^*,d^*}(t), \quad \forall (k,d) \neq (k^*,d^*).
\end{equation}

\noindent This strict inequality ensures that the global minimum of the cost function occurs only at the true fault location and driver type. For the blending-based approach, uniqueness is preserved through the continuous and monotonic relationship between the effective platoon length $N_{\rm eff}$ and the weights $W_1$ and $W_2$.

The comprehensive mathematical proof of these monotonicity properties, including the detailed analysis for both control architectures and the blending approach, is presented in Appendix~\ref{appendix:uniqueness_proof}.

\section{Simulation Results}
\begin{figure}[!t]
    \centering
    \subfloat[]{\includegraphics[width=0.49\columnwidth]{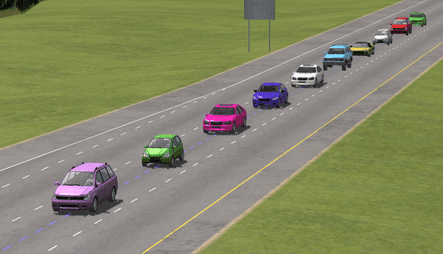}}
    \hfil
    \subfloat[]{\includegraphics[width=0.49\columnwidth]{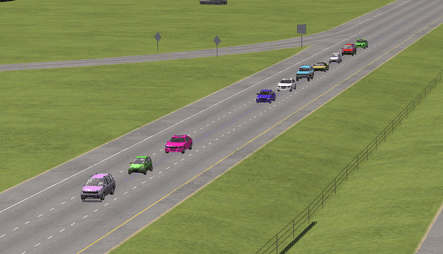}}
    \caption{(a) Platoon of Vehicles in CarSim (b) Platoon after Fault Occurrence.}
    \label{fig:carsim_combined}
\end{figure}

The high-fidelity CarSim/Simulink environment (Version 2022.0) was utilized to evaluate comprehensive vehicle maneuvers under various fault scenarios, as illustrated in Fig.~\ref{fig:carsim_combined}. The test scenarios included acceleration, constant-speed cruising, and braking conditions with fault injection in each case. The simulation utilized vehicle models from CarSim's pre-built library, incorporating detailed dynamic characteristics. The key specifications of the simulated vehicles are summarized in Table~\ref{tbl5}.

\begin{table}[!t]
\caption{Vehicle Parameters Value}
\label{tbl5}
\centering
\begin{tabular}{c c l}
\hline
Symbol & Quantity & Description \\
\hline
$m$ & 1530 kg & Vehicle mass \\
$I_z$ & 2315.3 kgm\textsuperscript{2} & Mass moment of inertia \\
$I_\omega$ & 0.8 kgm\textsuperscript{2} & Wheel moment of inertia \\
$L$ & 2.78 m & Wheelbase \\
$t_f, t_r$ & 1.55 m & Front and rear track width \\
$R_{\rm eff}$ & 0.325 m & Wheel effective radius \\
$a$ & 1.11 m & CG distance to front axle \\
$b$ & 1.67 m & CG distance to rear axle \\
$C_f$ & 80400 N/rad & Front tire cornering stiffness \\
$C_r$ & 82700 N/rad & Rear tire cornering stiffness \\
\hline
\end{tabular}
\end{table}

\begin{figure}[!t]
	\centering
	\includegraphics[width=\columnwidth]{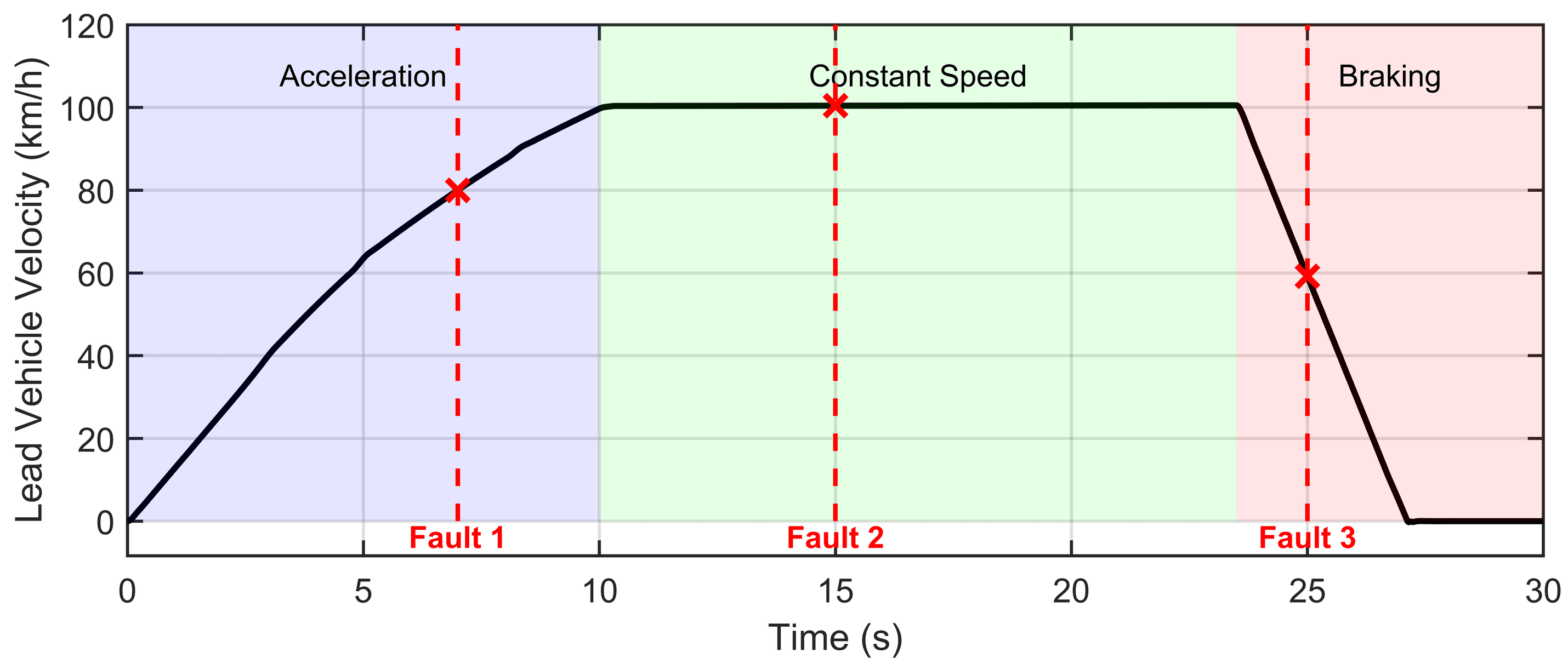}
	\caption{Lead Vehicle Velocity Profile with Fault Occurrences.}
	\label{Vehicle_Velocity}
\end{figure}

The fault occurrences are depicted in Fig.~\ref{Vehicle_Velocity}. In this study, a platoon of 10 vehicles is considered, with a fault occurring between vehicles 3 and 4, resulting in a remaining platoon length of N=7. Additionally, the driver taking control is considered to be distracted. For these scenarios, the tail vehicle displacement responses for all configurations are compared to the measured value of the last vehicle, and the error between these signals is calculated. Based on these measurements, the corresponding cost functions $J_{k,d}(t)$ are calculated according to equation \eqref{eq:cost_function}. The cost function parameters were selected as follows: $\alpha = 0.6$ to weigh the instantaneous error, $\beta = 0.4$ to account for historical error contribution, and $\lambda = 0.1$ as the forgetting factor to determine the influence of past errors.

\begin{figure}[!t]
	\centering
	\includegraphics[width=\columnwidth]{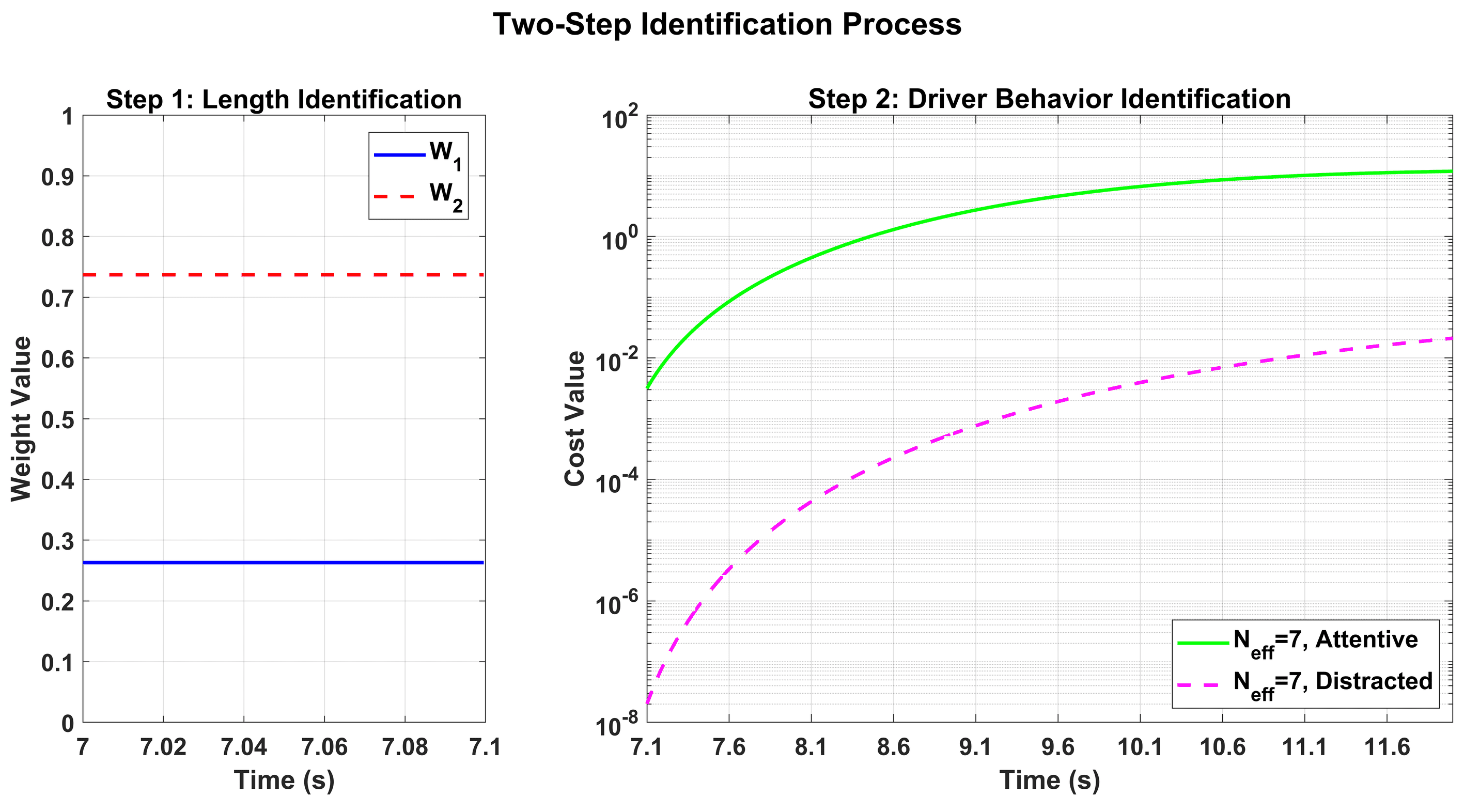}
	\caption{Simulation Results of Blending-Based Fault Detection Approach.}
	\label{fig:blending_weights}
\end{figure}

\begin{figure*}[!t]
    \centering
    \includegraphics[width=\textwidth]{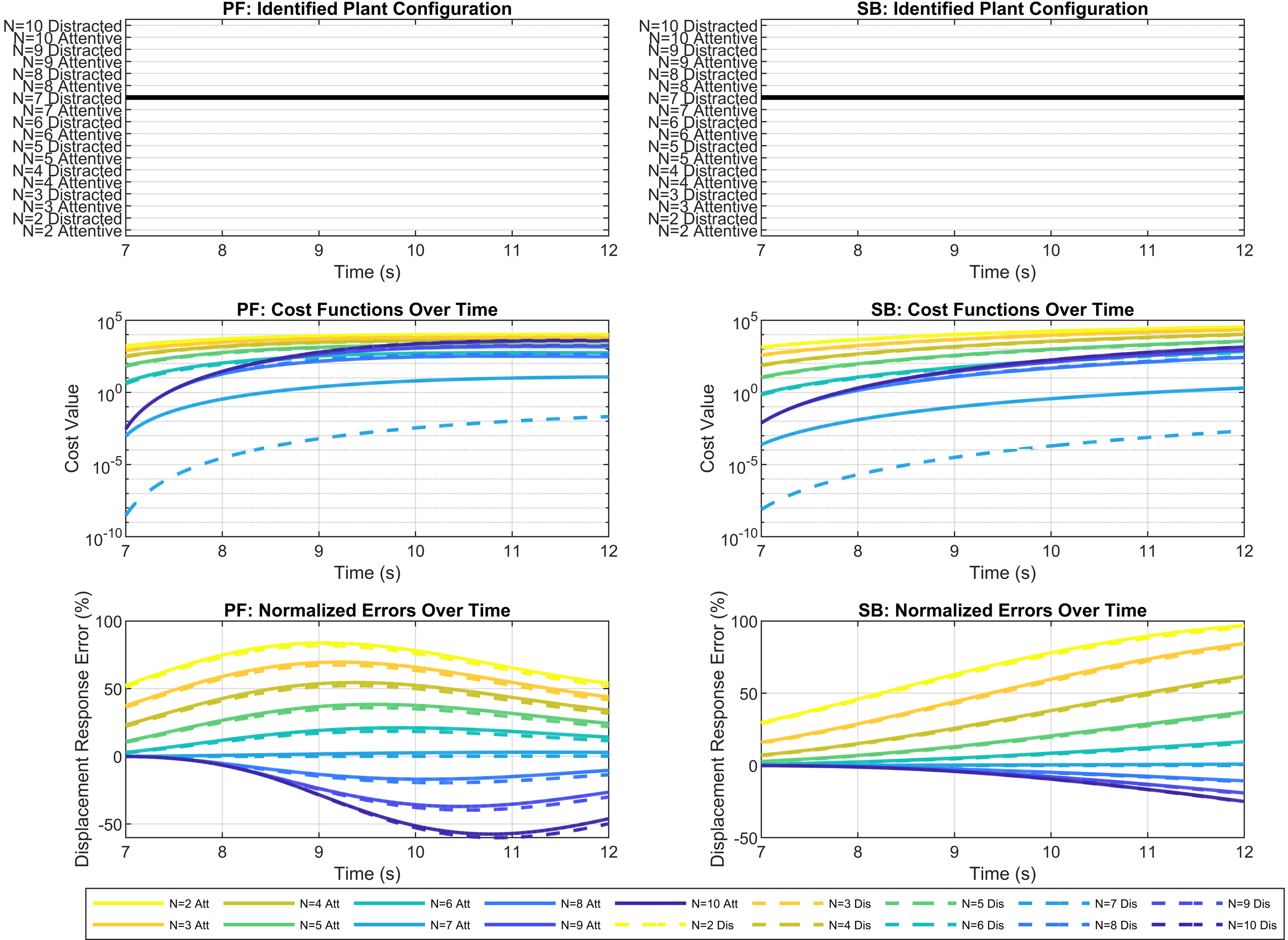}
    \caption{Scenario 1: Signal Drop at Acceleration: (a) Identified Configuration (b) Cost Value (c) Normalized Response Error.}
    \label{Results_S1}
\end{figure*}

\begin{figure*}[!t]
    \centering
    \includegraphics[width=\textwidth]{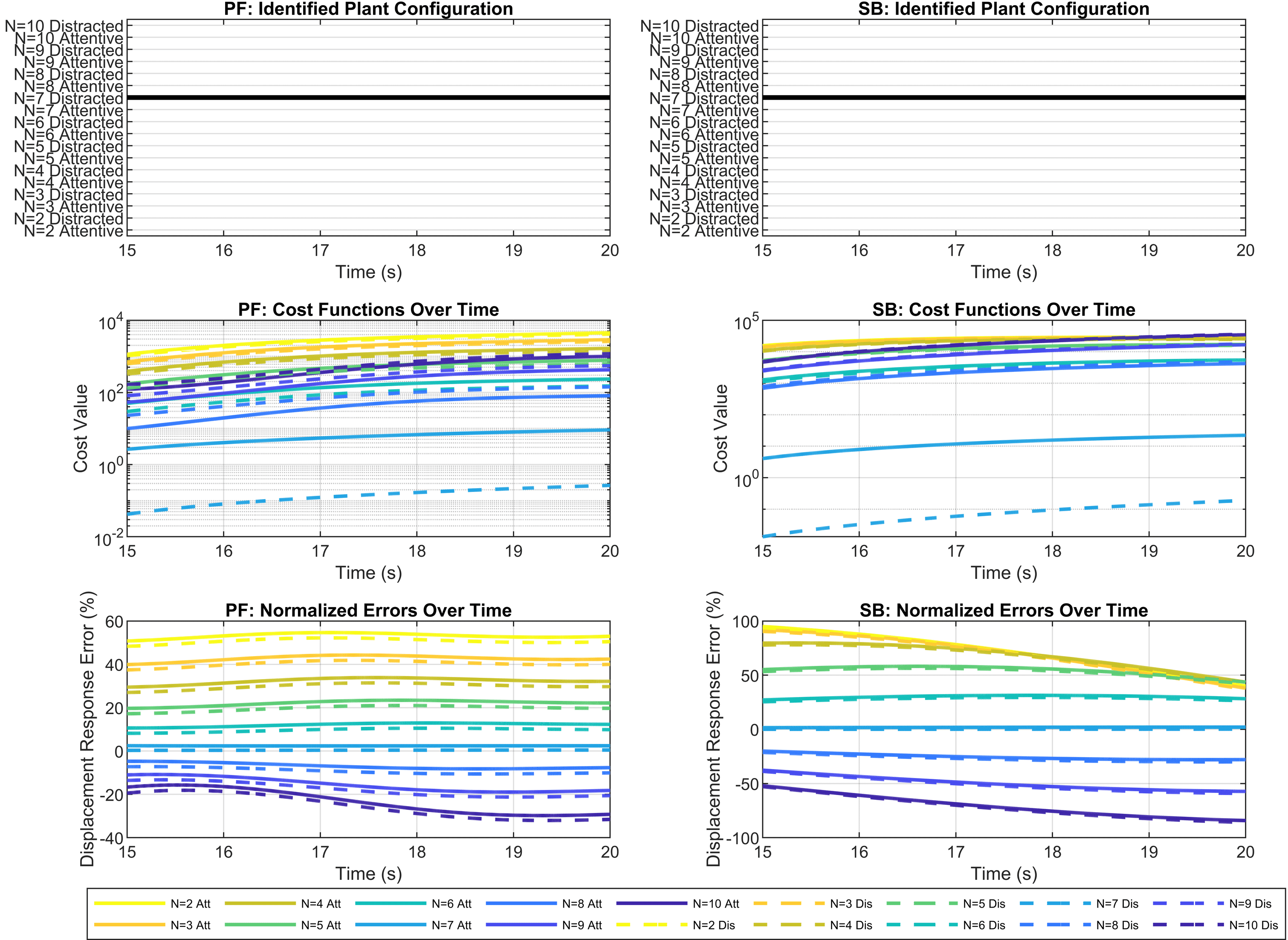}
    \caption{Scenario 2: Signal Drop at Constant Speed: (a) Identified Configuration (b) Cost Value (c) Normalized Response Error.}
    \label{Results_S2}
\end{figure*}

\begin{figure*}[!t]
    \centering
    \includegraphics[width=\textwidth]{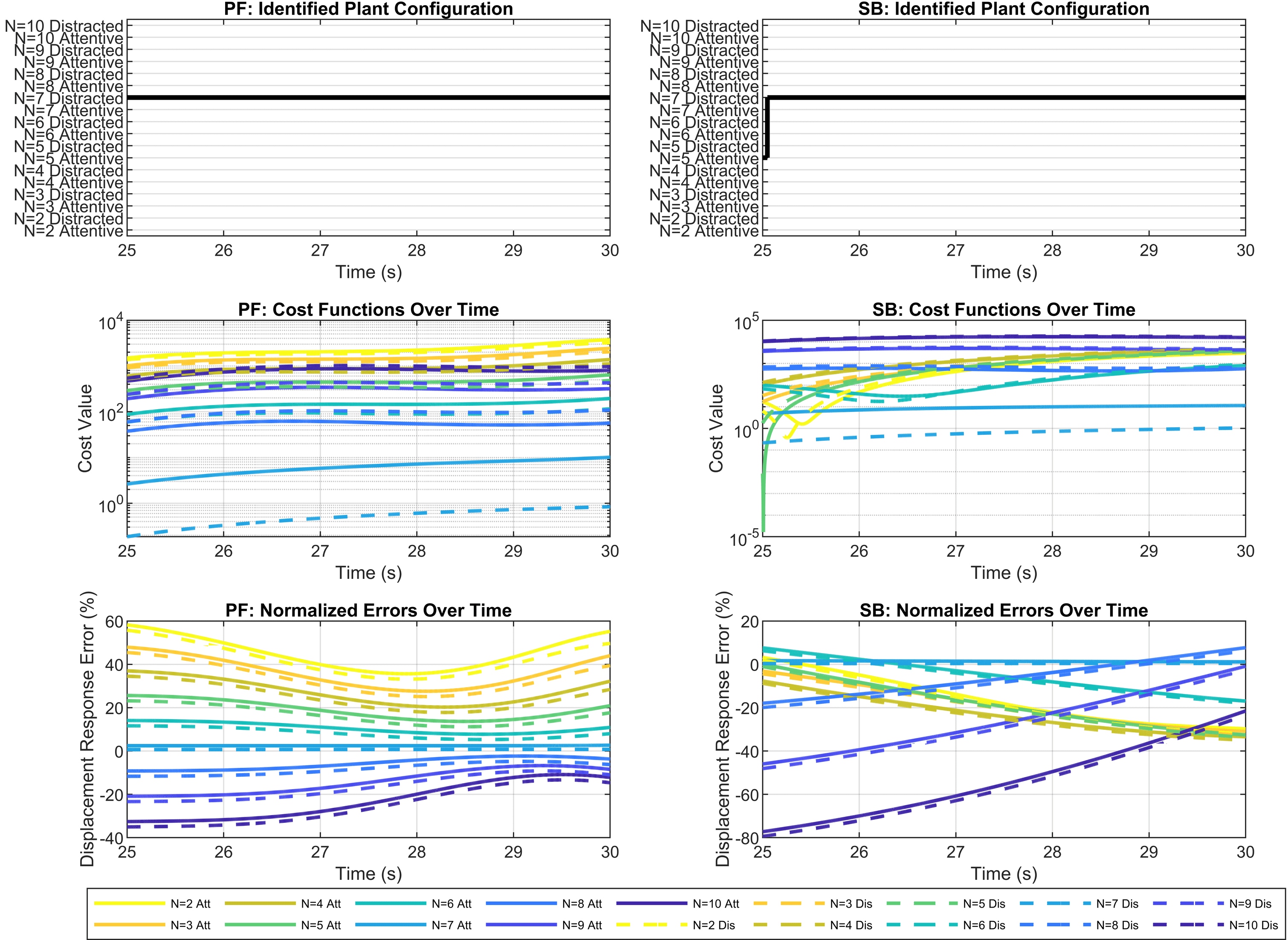}
    \caption{Scenario 3: Signal Drop at Braking: (a) Identified Configuration (b) Cost Value (c) Normalized Response Error.}
    \label{Results_S3}
\end{figure*}

Figs.~\ref{Results_S1} through~\ref{Results_S3} illustrate the identified system configurations for all three scenarios (signal drop during acceleration, constant speed, and braking) for both predecessor-following and symmetric bi-directional architectures, along with their corresponding cost values and normalized response errors over time. The simulation results demonstrate successful configuration identification across all scenarios. However, as shown in Fig.~\ref{Results_S3} for symmetric bi-directional architecture, the initial identification deviates from the true system configuration during the braking scenario, though it converges to the correct configuration after a brief period. This temporary misidentification can be attributed to the rapid variations in displacement error signals during braking, which increases the complexity of the identification process. To address this challenge in scenarios with sudden signal variations, one potential solution is to modify the learning parameters by increasing the coefficient $\beta$, which accounts for historical error, while decreasing $\alpha$, which corresponds to instantaneous error.

To demonstrate the effectiveness of the blending approach described in Section 3.4, we revisit the acceleration maneuver case with signal drop under the predecessor-following architecture, this time applying our computational reduction method. For the first step of identification, boundary models with platoon lengths of $N_1=2$ and $N_2=10$ are assigned weights $W_1$ and $W_2$ respectively. The evolution of these weights can be observed in Fig.~\ref{fig:blending_weights}. As shown, the weights are extracted using least squares optimization over a short time window (approximately 100 ms). From these weights and utilizing equation~\eqref{eq:blend}, we can determine $N_{\rm eff}$. In the subsequent identification step, driver behavior for this specific platoon length is analyzed. The configuration yielding the minimum error cost value is selected as the true configuration; as demonstrated in Fig.~\ref{fig:blending_weights}, the distracted driver model is identified. These results are consistent with our observations in Fig.~\ref{Results_S1}. Importantly, this blending approach successfully reduces the number of models required for identification, thereby decreasing computational complexity during real-time applications.

\section{Conclusion}
This paper presented a novel multi-model framework for simultaneous identification of signal drops and driver behavior in vehicular platoons using only tail vehicle measurements. Our approach successfully addresses two critical challenges in platoon systems: detecting communication failures and identifying driver attention states during human-in-loop scenarios.

The key contributions include a mathematical framework combining vehicle dynamics and driver models to characterize disturbance propagation, and a computationally efficient two-step blending approach using boundary models and weighted interpolation. This method significantly reduces computational complexity while maintaining accuracy, making it suitable for real-time applications. Simulation results using the CarSim/Simulink environment validated our approach across various operating conditions, demonstrating successful identification of both fault location and driver behavior type.

For future research, the investigation of more sophisticated fault scenarios including adversarial actions remains a promising direction.

\appendices
\section{Derivation of Transfer Functions}
\label{appendix:transfer_derivation}

This appendix presents the detailed derivation of the transfer functions for both predecessor-following and symmetric bidirectional architectures.

\subsection{Predecessor-Following Transfer Function Derivation}

Starting with the system dynamics, we take the Laplace transform of the error dynamics:
\begin{equation} \label{eq7}
    s^2\tilde{P}_i(s) = U_i(s) + W_i(s).
\end{equation}

\noindent The control law in the Laplace domain is:
\begin{equation} \label{eq8}
    U_i(s) = k_0(\tilde{P}_i(s) - \tilde{P}_{i-1}(s)) - b_0s(\tilde{P}_i(s) - \tilde{P}_{i-1}(s)).
\end{equation}

\noindent Substituting the control law and rearranging terms:
\begin{equation} \label{eq9}
    s^2\tilde{P}_i(s) = k_0(\tilde{P}_i(s) - \tilde{P}_{i-1}(s)) - b_0s(\tilde{P}_i(s) - \tilde{P}_{i-1}(s)) + W_i(s).
\end{equation}

\begin{equation} \label{eq10}
    (s^2 + b_0s + k_0)\tilde{P}_i(s) = (b_0s + k_0)\tilde{P}_{i-1}(s) + W_i(s).
\end{equation}

\noindent Each vehicle responds to both its predecessor's error signal and any disturbance it experiences:
\begin{equation} \label{eq11}
    \tilde{P}_i(s) = T(s) \tilde{P}_{i-1}(s) + \frac{W_i(s)}{s^2 + b_0 s + k_0},
\end{equation}

\noindent where:
\begin{equation} \label{eq12}
    T(s) = \frac{b_0 s + k_0}{s^2 + b_0 s + k_0}.
\end{equation}

\noindent Setting $W_i(s) = 0$ for $i = 1,\ldots,N$ yields:
\begin{equation} \label{eq13}
    \tilde{P}_i(s) = T(s) \tilde{P}_{i-1}(s).
\end{equation}

\noindent This recursive relationship leads to the final transfer function presented in equation \eqref{eq14}.

\subsection{Symmetric Bidirectional Transfer Function Derivation}

For the bidirectional case, the control law in the Laplace domain becomes:
\begin{equation} \label{eq15}
\begin{split}
    U_i(s) = & k_0(\tilde{P}_i(s) - \tilde{P}_{i-1}(s)) - b_0s(\tilde{P}_i(s) - \tilde{P}_{i-1}(s)) \\
    & - k_0(\tilde{P}_i(s) - \tilde{P}_{i+1}(s)) - b_0s(\tilde{P}_i(s) - \tilde{P}_{i+1}(s)).
\end{split}
\end{equation}

\noindent The system equations can be expressed in matrix form:
\begin{equation} \label{eq16}
\left[
\begin{smallmatrix}
    A_{11} & A_{12} & 0 & \cdots & 0 \\
    A_{21} & A_{22} & A_{23} & \cdots & 0 \\
    \vdots & \ddots & \ddots & \ddots & \vdots \\
    0 & \cdots & A_{n-1,n-2} & A_{n-1,n-1} & A_{n-1,n} \\
    0 & \cdots & 0 & A_{n,n-1} & A_{nn}
\end{smallmatrix}
\right]
\left[
\begin{smallmatrix}
    \tilde{P}_1(s) \\
    \tilde{P}_2(s) \\
    \vdots \\
    \tilde{P}_{N-1}(s) \\
    \tilde{P}_N(s)
\end{smallmatrix}
\right] = 
\left[
\begin{smallmatrix}
    0 \\
    0 \\
    \vdots \\
    0 \\
    0
\end{smallmatrix}
\right]
\end{equation}

\noindent with matrix elements:
\begin{align*}
    A_{ii} &= s^2 + 2b_0s + 2k_0 && \text{for } i = 1,\ldots,N-1 \\
    A_{nn} &= s^2 + b_0s + k_0 && \text{for } i = N \\
    A_{i,i+1} = A_{i+1,i} &= -(b_0s + k_0) && \text{for } i = 1,\ldots,N-1 \\
    A_{ij} &= 0 && \text{otherwise}
\end{align*}

\noindent Define key terms:
\begin{equation} \label{eq17}
\alpha(s) = s^2 + 2b_0s + 2k_0 \quad \text{(diagonal elements except last)},
\end{equation}
\begin{equation} \label{eq18}
\beta(s) = -(b_0s + k_0) \quad \text{(off-diagonal elements)},
\end{equation}
\begin{equation} \label{eq19}
\gamma(s) = s^2 + b_0s + k_0 \quad \text{(last diagonal element)}.
\end{equation}

\noindent Let $D_n(s)$ be the determinant of the $n\times n$ leading principal submatrix:
\begin{equation} \label{eq20}
D_n(s) = \alpha(s)D_{n-1}(s) - \beta(s)^2D_{n-2}(s).
\end{equation}

\noindent For $n = N$:
\begin{equation} \label{eq21}
D_N(s) = \gamma(s)D_{N-1}(s) - \beta(s)^2D_{N-2}(s).
\end{equation}

\noindent With initial conditions:
\begin{equation} \label{eq22}
D_1(s) = \alpha(s) = s^2 + 2b_0s + 2k_0,
\end{equation}
\begin{equation} \label{eq23}
D_2(s) = \alpha(s)^2 - \beta(s)^2 = (s^2 + 2b_0s + 2k_0)^2 - (b_0s + k_0)^2.
\end{equation}

\noindent These recursive relationships lead to the final transfer function presented in equations \eqref{eq24} and \eqref{eq25}.

\section{Derivation of Effective Platoon Length}
\label{appendix:blending_derivation}

This appendix presents the comprehensive mathematical derivation of the effective platoon length $N_{\rm eff}$ from the transfer function weights in the blending multi-model approach.

\subsection{Problem Statement and Initial Conditions}

Given a weighted sum of transfer functions representing the predicted system behavior:
\begin{equation}
    G_{\rm pred}(s) = W_1G_1(s) + W_2G_2(s),
\end{equation}

\noindent where the component transfer functions are defined as:
\begin{equation}
    G_1(s) = \left(\frac{b_0s + k_0}{s^2 + b_0s + k_0}\right)^{N_1-1} H(s),
\end{equation}
\begin{equation}
    G_2(s) = \left(\frac{b_0s + k_0}{s^2 + b_0s + k_0}\right)^{N_2-1} H(s),
\end{equation}

\noindent with the fundamental assumption:
\begin{equation}
    N_2 > N_1,
\end{equation}

\noindent where $N_1$ and $N_2$ span all possible platoon lengths that may occur after signal drops and connection failures in the system.

The weights must satisfy:
\begin{align}
    W_1 + W_2 &= 1, \\
    W_1, W_2 &\geq 0.
\end{align}

The objective is to find $N_{\rm eff}$ such that:
\begin{equation}
    G_{\rm pred}(s) = \left(\frac{b_0s + k_0}{s^2 + b_0s + k_0}\right)^{N_{\rm eff}-1} H(s).
\end{equation}

\subsection{Derivation Steps}

\noindent I. Base Transfer Function Definition
\begin{equation}
    G_{\rm base}(s) = \frac{b_0s + k_0}{s^2 + b_0s + k_0}.
\end{equation}

\noindent II. Initial Equation Transformation\\
The original equation becomes:
\begin{equation}
    \begin{split}
    W_1[G_{\rm base}(s)]^{N_1-1}H(s) + W_2[G_{\rm base}(s)]^{N_2-1}H(s) \\
    = [G_{\rm base}(s)]^{N_{\rm eff}-1}H(s).
    \end{split}
\end{equation}

\noindent III. Factor Common Terms\\
Factor out $H(s)$:
\begin{equation}
    \begin{split}
    H(s)\left(W_1[G_{\rm base}(s)]^{N_1-1} + W_2[G_{\rm base}(s)]^{N_2-1}\right) \\
    = H(s)[G_{\rm base}(s)]^{N_{\rm eff}-1}.
    \end{split}
\end{equation}

\noindent Since this must be true for all $s$, and canceling $H(s)$:
\begin{equation}
    W_1[G_{\rm base}(s)]^{N_1-1} + W_2[G_{\rm base}(s)]^{N_2-1} = [G_{\rm base}(s)]^{N_{\rm eff}-1}.
\end{equation}

\noindent IV. Variable Substitution\\
Let $x = G_{\rm base}(s)$. Then for all $x$:
\begin{equation}
    W_1x^{N_1-1} + W_2x^{N_2-1} = x^{N_{\rm eff}-1}.
\end{equation}

\noindent V. Logarithmic Transformation\\
Take the natural logarithm of both sides:
\begin{equation}
    \ln(W_1x^{N_1-1} + W_2x^{N_2-1}) = (N_{\rm eff}-1)\ln(x).
\end{equation}

\noindent Let $y = \ln(x)$. Then:
\begin{equation}
    \ln(W_1e^{(N_1-1)y} + W_2e^{(N_2-1)y}) = (N_{\rm eff}-1)y.
\end{equation}

\noindent VI. Analysis of Exponential Terms\\
Define the ratio $R$ of the exponential terms:
\begin{equation}
    R = \frac{W_1e^{(N_1-1)y}}{W_2e^{(N_2-1)y}} = \frac{W_1}{W_2}e^{(N_1-N_2)y}.
\end{equation}

\noindent At the critical point where both terms contribute equally, $R = 1$:
\begin{equation}
    \frac{W_1}{W_2}e^{(N_1-N_2)y_c} = 1.
\end{equation}

\noindent This critical point has the following characteristics:
\begin{itemize}
    \item At $y_c$, $W_1e^{(N_1-1)y_c} = W_2e^{(N_2-1)y_c}$
    \item For $y < y_c$: The term $W_1e^{(N_1-1)y}$ dominates
    \item For $y > y_c$: The term $W_2e^{(N_2-1)y}$ dominates
\end{itemize}

\noindent The uniqueness of this point is guaranteed because:
\begin{itemize}
    \item The function $R$ is strictly monotonic in $y$
    \item It decreases from $\infty$ to $0$ as $y$ goes from $-\infty$ to $\infty$ (when $W_1, W_2 > 0$)
    \item Therefore, exactly one point exists where $R = 1$
\end{itemize}

\noindent VII. Critical Point Solution\\
Solve for the critical point $y_c$:
\begin{equation}
    y_c = \frac{1}{N_2-N_1}\ln\left(\frac{W_1}{W_2}\right).
\end{equation}

\noindent VIII. Substitution and Expansion\\
Substitute $y_c$ back into the original equation:
\begin{equation}
    W_1e^{(N_1-1)y_c} + W_2e^{(N_2-1)y_c} = e^{(N_{\rm eff}-1)y_c}.
\end{equation}

\noindent Expand using $y_c$:
\begin{equation}
    \begin{split}
    W_1e^{(N_1-1)\ln(W_1/W_2)/(N_2-N_1)} + W_2e^{(N_2-1)\ln(W_1/W_2)/(N_2-N_1)} \\
    = e^{(N_{\rm eff}-1)\ln(W_1/W_2)/(N_2-N_1)}.
    \end{split}
\end{equation}

\noindent IX. Final Solution\\
Simplify using properties of exponentials:
\begin{equation}
    \begin{split}
    W_1(W_1/W_2)^{(N_1-1)/(N_2-N_1)} + W_2(W_1/W_2)^{(N_2-1)/(N_2-N_1)} \\
    = (W_1/W_2)^{(N_{\rm eff}-1)/(N_2-N_1)}.
    \end{split}
\end{equation}

\noindent Take natural logarithm of both sides and solve for $N_{\rm eff}$:
\begin{equation}
    \begin{split}
    N_{\rm eff} = 1 + (N_2-N_1) & \frac{\ln\big(W_1(W_1/W_2)^{(N_1-1)/(N_2-N_1)} + }{} \\
    & \frac{W_2(W_1/W_2)^{(N_2-1)/(N_2-N_1)}\big)}{\ln(W_1/W_2)}.
    \end{split}
\end{equation}

\subsection{Implementation Considerations}

\noindent I. Numerical Stability\\
For numerical stability, we need to consider cases where:
\begin{enumerate}
    \item $W_1$ or $W_2$ are very close to 0
    \item $W_1/W_2$ is very close to 1
    \item $N_2-N_1$ is very small
\end{enumerate}

\noindent II. Integer Conversion\\
Since $N_{\rm eff}$ must be an integer and the formula gives a continuous value, we have different options:

\begin{enumerate}
    \item Direct Rounding:
    \begin{equation}
        N_{\rm fin} = \text{round}(N_{\rm eff}).
    \end{equation}

    \item Error Minimization:
    \begin{align}
        N_{\rm fin} &= \operatorname{argmin}_{N \in S} \|G_N(s) - G_{\rm pred}(s)\|, \\
        S &= \{\lfloor N_{\rm eff} \rfloor, \lceil N_{\rm eff} \rceil\}.
    \end{align}
\end{enumerate}

\section{Proof of Identification Solution Uniqueness}
\label{appendix:uniqueness_proof}

This appendix provides the complete mathematical analysis establishing the uniqueness of the fault location and driver behavior identification solution in the proposed multi-model framework.

\subsection{Cost Function Analysis}

The identification process seeks to minimize the cost function:

\begin{equation}
    J_{k,d}(t) = \alpha\|e_{k,d}(t)\|^2 + \beta \int_{t_f}^t e^{-\lambda(t-\tau)}\|e_{k,d}(\tau)\|^2 d\tau,
\end{equation}

\noindent where $e_{k,d}(t) = y_N(t) - \hat{y}_{k,d}(t)$ represents the identification error between the measured output from the last vehicle and the predicted output from model $M_{k,d}$.

To establish uniqueness, we must prove that the cost function has a unique global minimum at the true fault location $k^*$ and the true driver type $d^*$.

\subsection{Monotonicity Analysis for Predecessor-Following Architecture}

For the predecessor-following architecture, the transfer function from the first to the last vehicle is:

\begin{equation}
    G_{\rm PF}(s) = \frac{\tilde{P}_N(s)}{\tilde{P}_1(s)}= \left(\frac{b_0 s + k_0}{s^2 + b_0 s + k_0}\right)^{N-1}.
\end{equation}

\noindent When a fault occurs at position $k$, the effective transfer function becomes:

\begin{equation}
    G_{k,d}(s) = \left(\frac{b_0 s + k_0}{s^2 + b_0 s + k_0}\right)^{N-k} \cdot H_d(s),
\end{equation}

\noindent where $H_d(s)$ represents the driver model transfer function (either attentive or distracted).

To establish monotonicity, we analyze the gradient of the cost function with respect to the length of the remaining platoon $(N-k)$:

\begin{equation}
\begin{split}
    \frac{\partial J_{k,d}(t)}{\partial (N-k)} &= 2\alpha\langle e_{k,d}(t), \frac{\partial e_{k,d}(t)}{\partial (N-k)}\rangle \\
    &+ 2\beta\int_{t_f}^t e^{-\lambda(t-\tau)}\langle e_{k,d}(\tau), \frac{\partial e_{k,d}(\tau)}{\partial (N-k)}\rangle d\tau.
\end{split}
\end{equation}

\subsubsection{Derivative Analysis}

The partial derivative of the error with respect to the remaining platoon length is:

\begin{equation}
    \frac{\partial e_{k,d}(t)}{\partial (N-k)} = -\frac{\partial \hat{y}_{k,d}(t)}{\partial (N-k)} = -\frac{\partial}{\partial (N-k)}\mathcal{L}^{-1}\left\{G_{k,d}(s)U(s)\right\}.
\end{equation}

\noindent Let $T(s) = \frac{b_0 s + k_0}{s^2 + b_0 s + k_0}$. The model output can be expressed as:

\begin{equation}
    \hat{y}_{k,d}(t) = \mathcal{L}^{-1}\left\{T(s)^{N-k} \cdot H_d(s) \cdot U(s)\right\}.
\end{equation}

\noindent Taking the derivative with respect to $(N-k)$:

\begin{equation}
\begin{split}
    \frac{\partial \hat{y}_{k,d}(t)}{\partial (N-k)} &= \mathcal{L}^{-1}\left\{ \frac{\partial T(s)^{N-k}}{\partial (N-k)} \cdot H_d(s) \cdot U(s) \right\} \\
    &= \mathcal{L}^{-1}\left\{ T(s)^{N-k} \cdot \ln(T(s)) \cdot H_d(s) \cdot U(s) \right\}.
\end{split}
\end{equation}

\subsubsection{Sign Analysis}

For the predecessor-following architecture with typical string stable parameters, $|T(j\omega)| < 1$ for $\omega > 0$. This implies that:

\begin{equation}
    \Re\{\ln(T(j\omega))\} < 0.
\end{equation}

\noindent Consequently, when $k < k^*$ (overestimating the remaining platoon length), the model output will be smaller than the true output, making $e_{k,d}(t) > 0$. Since $\frac{\partial \hat{y}_{k,d}(t)}{\partial (N-k)} < 0$, we have $\frac{\partial e_{k,d}(t)}{\partial (N-k)} > 0$ and thus $\frac{\partial J_{k,d}(t)}{\partial (N-k)} > 0$.

Conversely, when $k > k^*$ (underestimating the remaining platoon length), the model output will be larger than the true output, making $e_{k,d}(t) < 0$. This leads to $\frac{\partial J_{k,d}(t)}{\partial (N-k)} < 0$.

This sign pattern establishes that the cost function $J_{k,d}(t)$ has a unique minimum at the true fault location $k^*$ when the correct driver type $d^*$ is selected.

\subsection{Analytical Proof for Symmetric Bidirectional Architecture}

For the symmetric bidirectional architecture, the transfer function from the first to the last vehicle is:

\begin{equation}
    G_{\rm SB}(s) = \frac{\tilde{P}_N(s)}{\tilde{P}_1(s)} = \frac{(-1)^{N+1}\beta(s)^{N-1}}{D_N(s)},
\end{equation}

\noindent where $\beta(s) = -(b_0s + k_0)$ and $D_N(s)$ is the denominator polynomial defined in equation \eqref{eq25}. When a fault occurs at position $k$ with driver behavior $d$, the effective transfer function becomes:

\begin{equation}
    G_{k,d}(s) = \frac{(-1)^{N-k+2}\beta(s)^{N-k}}{D_{N-k+1}(s)} \cdot H_d(s),
\end{equation}

\noindent where $H_d(s)$ represents the driver model (either attentive or distracted).

\subsubsection{Structural Distinctness of Transfer Functions}

To establish uniqueness of the identification solution, we must prove that for any $(k_1,d_1) \neq (k_2,d_2)$, the transfer functions $G_{k_1,d_1}(s)$ and $G_{k_2,d_2}(s)$ cannot produce identical outputs for all inputs.

First, consider the case where $d_1 = d_2 = d$ but $k_1 \neq k_2$. The difference between these transfer functions is:

\begin{equation}
\begin{split}
    \Delta G(s) &= G_{k_1,d}(s) - G_{k_2,d}(s) \\
    &= \Bigg[ \frac{(-1)^{N-k_1+2}\beta(s)^{N-k_1}}{D_{N-k_1+1}(s)} \\
    &\quad - \frac{(-1)^{N-k_2+2}\beta(s)^{N-k_2}}{D_{N-k_2+1}(s)} \Bigg] \cdot H_d(s).
\end{split}
\end{equation}

\noindent For $\Delta G(s)$ to be zero for all $s$, we would need:

\begin{equation}
    \frac{(-1)^{N-k_1+2}\beta(s)^{N-k_1}}{D_{N-k_1+1}(s)} = \frac{(-1)^{N-k_2+2}\beta(s)^{N-k_2}}{D_{N-k_2+1}(s)}.
\end{equation}

\noindent Without loss of generality, assume $k_1 < k_2$. After simplification, this yields:

\begin{equation}
    \beta(s)^{k_2-k_1}D_{N-k_2+1}(s) = D_{N-k_1+1}(s).
\end{equation}

\noindent This is a polynomial equality that must hold for all $s$. However, the degrees of these polynomials are:

\begin{itemize}
    \item Left side: $(k_2-k_1) + 2(N-k_2+1) = 2N - k_2 + k_2 - k_1 + 2 = 2N - k_1 + 2$
    \item Right side: $2(N-k_1+1) = 2N - 2k_1 + 2$
\end{itemize}

\noindent Since $k_1 < k_2$, these degrees differ, creating a contradiction. Therefore, transfer functions with different fault locations must be distinct.

\subsubsection{Recurrence Relation Analysis}

The denominator polynomials follow the recurrence relation:

\begin{equation}
    D_n(s) = \alpha(s)D_{n-1}(s) - \beta(s)^2D_{n-2}(s),
\end{equation}

\noindent where $\alpha(s) = s^2 + 2b_0s + 2k_0$. This recurrence relation generates polynomials with distinct root patterns for each value of $n$. The roots of $D_n(s)$ correspond to the poles of the transfer function $G_{k,d}(s)$.

For a platoon with fault at position $k$, these poles are distributed as:

\begin{equation}
\begin{split}
    p_{k,i} &= -b_0 \pm j\sqrt{2k_0 - \frac{b_0^2}{2}} \cdot \sin\left(\frac{i\pi}{2(N-k+1)}\right), \\
    &\quad \text{for } i=1,2,\ldots,(N-k+1).
\end{split}
\end{equation}

\noindent For any two different fault locations $k_1$ and $k_2$, the pole patterns differ in:
\begin{enumerate}
    \item The number of poles ($(N-k_1+1)$ vs. $(N-k_2+1)$)
    \item The spacing of the imaginary parts through the $\sin(\frac{i\pi}{2(N-k+1)})$ term
\end{enumerate}

\subsubsection{Minimum Separation Property}

Define $\rho(k_1, k_2)$ as the minimum distance between any pole from the $k_1$ configuration and any pole from the $k_2$ configuration:

\begin{equation}
    \rho(k_1, k_2) = \min_{i,j} |p_{k_1,i} - p_{k_2,j}|.
\end{equation}

\noindent Due to the structure of the sine function in the pole expressions, we can establish:

\begin{equation}
    \rho(k_1, k_2) \geq C \cdot |k_1 - k_2|,
\end{equation}

\noindent for some positive constant $C$. This minimum separation property ensures that the difference between transfer functions grows at least linearly with the difference in fault locations.

\subsubsection{Driver Type Distinctness}

For the case where $k_1 = k_2 = k$ but $d_1 \neq d_2$, the difference between transfer functions is:

\begin{equation}
\begin{split}
    \Delta G(s) &= G_{k,d_1}(s) - G_{k,d_2}(s) \\
    &= \frac{(-1)^{N-k+2}\beta(s)^{N-k}}{D_{N-k+1}(s)} \cdot [H_{d_1}(s) - H_{d_2}(s)].
\end{split}
\end{equation}

\noindent The driver transfer functions have fundamentally different structures:

\begin{align}
    H_{\rm att}(s) &= \frac{1 + 5.41s}{1 + 2(0.54)(4.15)s + (4.15)^2 s^2} e^{-0.324s},\\
    H_{\rm dist}(s) &= \frac{1 + 6.96s}{1 + 2(0.65)(4.76)s + (4.76)^2 s^2} e^{-0.512s}.
\end{align}

\noindent The different time delays (0.324s vs. 0.512s) and transfer function coefficients ensure that $H_{d_1}(s) - H_{d_2}(s) \neq 0$ for almost all values of $s$, guaranteeing distinct responses.

\subsubsection{Combined Uniqueness Result}

The analysis above establishes that the transfer functions $G_{k_1,d_1}(s)$ and $G_{k_2,d_2}(s)$ are distinct whenever $(k_1,d_1) \neq (k_2,d_2)$. This ensures that the identification error $e_{k,d}(t) = y_N(t) - \hat{y}_{k,d}(t)$ is minimized only when $(k,d) = (k^*,d^*)$, where $(k^*,d^*)$ represents the true fault location and driver type.

Consequently, the cost function $J_{k,d}(t)$ has a unique global minimum at $(k^*,d^*)$, establishing the uniqueness of the identification solution for the bidirectional architecture.

\subsection{Driver Behavior Differentiation}

For a fixed fault location $k$, the cost function differentiates between driver behavior types due to the distinct transfer functions of attentive and distracted drivers:

\begin{equation}
    \|H_{\rm att}(j\omega) - H_{\rm dist}(j\omega)\|_{\mathcal{H}_2} > \gamma_2,
\end{equation}

\noindent where $\gamma_2$ is a positive constant, and $H_{\rm att}(s)$ and $H_{\rm dist}(s)$ represent the transfer functions for attentive and distracted drivers, respectively.

This separation ensures that:

\begin{equation}
    \|e_{k,{\rm att}}(t) - e_{k,{\rm dist}}(t)\| > \gamma_3 > 0,
\end{equation}

\noindent for some positive threshold $\gamma_3$, allowing unique identification of both fault location and driver type.

\subsection{Uniqueness in Blending-Based Approach}

For the blending-based approach, uniqueness is established by proving three key properties:

\subsubsection{Monotonicity of Effective Platoon Length}

The effective platoon length $N_{\rm eff}$ varies monotonically with the weights $W_1$ and $W_2$:

\begin{equation}
    \frac{\partial N_{\rm eff}}{\partial W_1} > 0, \quad \frac{\partial N_{\rm eff}}{\partial W_2} < 0,
\end{equation}

\noindent when $N_2 > N_1$.

\subsubsection{Convexity of Error Surface}

The error surface formed by the cost function $J(W_1, W_2, d)$ is convex with respect to the weights:

\begin{equation}
    \frac{\partial^2 J}{\partial W_1^2} > 0, \quad \frac{\partial^2 J}{\partial W_2^2} > 0, \quad \frac{\partial^2 J}{\partial W_1 \partial W_2} < 0.
\end{equation}

\subsubsection{Convergence of Optimization}

Under sufficient excitation conditions, the least squares optimization for weights $W_1$ and $W_2$ converges to a unique solution that minimizes the identification error:

\begin{equation}
\begin{split}
    (W_1^*, W_2^*) = \arg\min_{W_1+W_2=1,\, W_1,W_2 \geq 0}\, &\|y_N(t) - W_1 \hat{y}_{1,d}(t) \\
    &- W_2 \hat{y}_{2,d}(t)\|^2.
\end{split}
\end{equation}

\noindent These properties together ensure that the blending-based approach preserves the uniqueness of the identification solution.

In summary, the comprehensive analysis presented in this appendix establishes that the proposed multi-model identification framework guarantees a unique solution for both fault location and driver type due to the inherent monotonicity properties of the cost function with respect to the length of the platoon behind the faulty vehicle.


\subsection*{} 
\setlength\intextsep{0pt} 
\begin{wrapfigure}{l}{0.13\textwidth}
    \centering
    \includegraphics[width=0.15\textwidth]{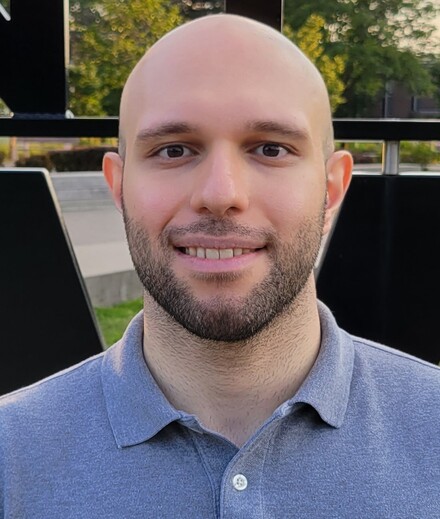}
\end{wrapfigure}
\noindent {\small \textbf{Farid Mafi} received the B.Sc. and M.Sc. degrees in mechanical engineering from the Sharif University of Technology, Tehran, Iran in 2017 and 2019, respectively. He is currently a Ph.D. Candidate at the University of Waterloo, specializing in vehicle control systems. His research interests include vehicle dynamics, control, estimation, and optimization.}

\subsection*{} 
\setlength\intextsep{0pt} 
\begin{wrapfigure}{l}{0.13\textwidth}
    \centering
    \includegraphics[width=0.15\textwidth]{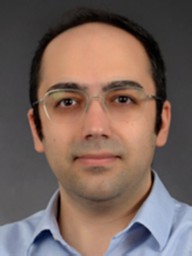}
\end{wrapfigure}
\noindent {\small \textbf{Mohammad Pirani} (Senior Member, IEEE) received the M.A.Sc. degree in electrical and computer engineering and a Ph.D. degree in mechanical and mechatronics engineering from the University of Waterloo, in 2014 and 2017, respectively. He is an Assistant Professor at the Department of Mechanical Engineering, University of Ottawa, Canada. He was a Research Assistant Professor with the Department of Mechanical and Mechatronics Engineering, University of Waterloo, from 2022 to 2023. He held post-doctoral researcher positions with the University of Toronto, from 2019 to 2021, and the KTH Royal Institute of Technology, Sweden, from 2018 to 2019. His research interests include resilient and fault-tolerant control, networked control systems, and multi-agent systems.}

\end{document}